\documentclass[twocolumn, superscriptaddress, nobibnotes]{revtex4-2}

\usepackage{graphicx}
\usepackage{physics}
\usepackage{amssymb}
\usepackage{amsmath}
\usepackage{color}
\usepackage{graphicx}
\usepackage{dcolumn}
\usepackage{epstopdf}
\usepackage{bm}
\usepackage{natbib}
\usepackage{tikz}
\usepackage[colorlinks=true,citecolor=blue]{hyperref}

\begin{document}

\title{Continuous microwave photon counting by semiconductor-superconductor hybrids}

\author{Subhomoy Haldar}
\email{subhomoy.haldar@ftf.lth.se}
 \address{NanoLund and Solid State Physics, Lund University, Box 118, 22100 Lund, Sweden}
   \author{David Barker}
 \address{NanoLund and Solid State Physics, Lund University, Box 118, 22100 Lund, Sweden}
  \author{Harald Havir}
 \address{NanoLund and Solid State Physics, Lund University, Box 118, 22100 Lund, Sweden}
   \author{Antti Ranni}
 \address{NanoLund and Solid State Physics, Lund University, Box 118, 22100 Lund, Sweden}
 \author{Sebastian Lehmann}
 \address{NanoLund and Solid State Physics, Lund University, Box 118, 22100 Lund, Sweden}
\author{Kimberly A. Dick }
 \address{NanoLund and Solid State Physics, Lund University, Box 118, 22100 Lund, Sweden}
 \address{Center for Analysis and Synthesis, Lund University, Box 124, 22100 Lund, Sweden}
 \author{Ville F. Maisi}
  \email{ville.maisi@ftf.lth.se}
 \address{NanoLund and Solid State Physics, Lund University, Box 118, 22100 Lund, Sweden}

\date{\today}
\begin{abstract}

The growing interest in quantum information has enabled the manipulation and readout of microwave photon states with high fidelities. The presently available microwave photon counters, based on superconducting circuits, are limited to non-continuous pulsed mode operation, requiring additional steps for qubit state preparation before an actual measurement. Here, we present a continuous microwave photon counter based on superconducting cavity-coupled semiconductor quantum dots. The device utilizes photon-assisted tunneling in a double quantum dot with tunneling events being probed by a third dot. Our device detects both single and multiple-photon absorption events independently, thanks to the energy tunability of a two-level double-dot absorber. We show that the photon-assisted tunnel rates serve as the measure of the cavity photon state in line with the $P(E)$ theory -- a theoretical framework delineating the mediation of the cavity photon field via a two-level environment. We further describe the single photon detection using the Jaynes-Cummings input-output theory and show that it agrees with the $P(E)$ theory predictions.

\end{abstract}
\maketitle
The field of quantum information science and quantum computing has seen significant advancements in recent years~\cite{bennett2000, scappucci2021, sigillito2019, yang2004, sebastian2013}. One crucial aspect here is the ability to read quantum states with high fidelity and minimal error rates~\cite{chao2019, Heinsoo2018, gonzalez2021}. In this domain, many leading approaches rely on their ability to measure and manipulate microwave signals in the circuit quantum electrodynamics architecture~\cite{naik2017, blais2020, Mario2019}. Extensive research has been performed to realize microwave photon counters based on superconducting circuits, enabling reliable and accurate information extraction~\cite{opremcak208,chen2011,opremcak2021, pankratov2022}. The microwave photon detection using superconducting circuits has also been demonstrated within a quantum non-demolition framework~\cite{johnson2010, besse2018, kono2018}. However, these developments primarily center around non-continuous pulsed-mode detection or require additional steps to prepare the qubit before an actual measurement. 

An alternative avenue lies in the superconductor cavity-coupled semiconductor double quantum dot (DQD)~\cite{burkard2020, chatterjee2021, frey2012}. Recent research has extensively investigated the interaction of cavity photons with DQDs, analyzing the results using the Jaynes-Cummings input-output theory~\cite{clerk2020, frey2012, basset2013, mi2017, burkard2020, scarlino2021, ranni2023}. A detailed understanding of this interaction between microwave photons and a semiconducting qubit has spurred the development of hybrid devices, demonstrating remarkable capabilities in continuous and highly efficient microwave photon-to-electron conversion~\cite{khan2021, wong2017, ghirri2020}. In this framework, a quantum point contact near a DQD absorber has also been used to map the charge stability diagram of the DQD and probe the signatures of photon-assisted tunneling~\cite{petta2004, gustavsson2007}. However, a microwave photon counter capable of operating continuously and probing single or multiple photon absorption events individually remains an unmet challenge.

Here, we present the first experimental realization of a microwave photon counter using superconductor-semiconductor hybrids, which is capable of operating continuously without requiring pulse control. We measure photon-assisted tunnel rates with varying microwave input powers and analyze the results within the environmental $P(E)$ theory framework with the Tien-Gordon coefficients~\cite{tien1963}. We show further that the single-photon detection efficiency of our device using the Jaynes-Cummings input-output theory matches the results of the $P(E)$ theory. With the capability to individually and continuously detect single and multiple photon absorption events, our findings pave a path to exploring photons and their statistics beyond the pulsed-control time.

\begin{figure*}[t]
\includegraphics[width=6.0in]{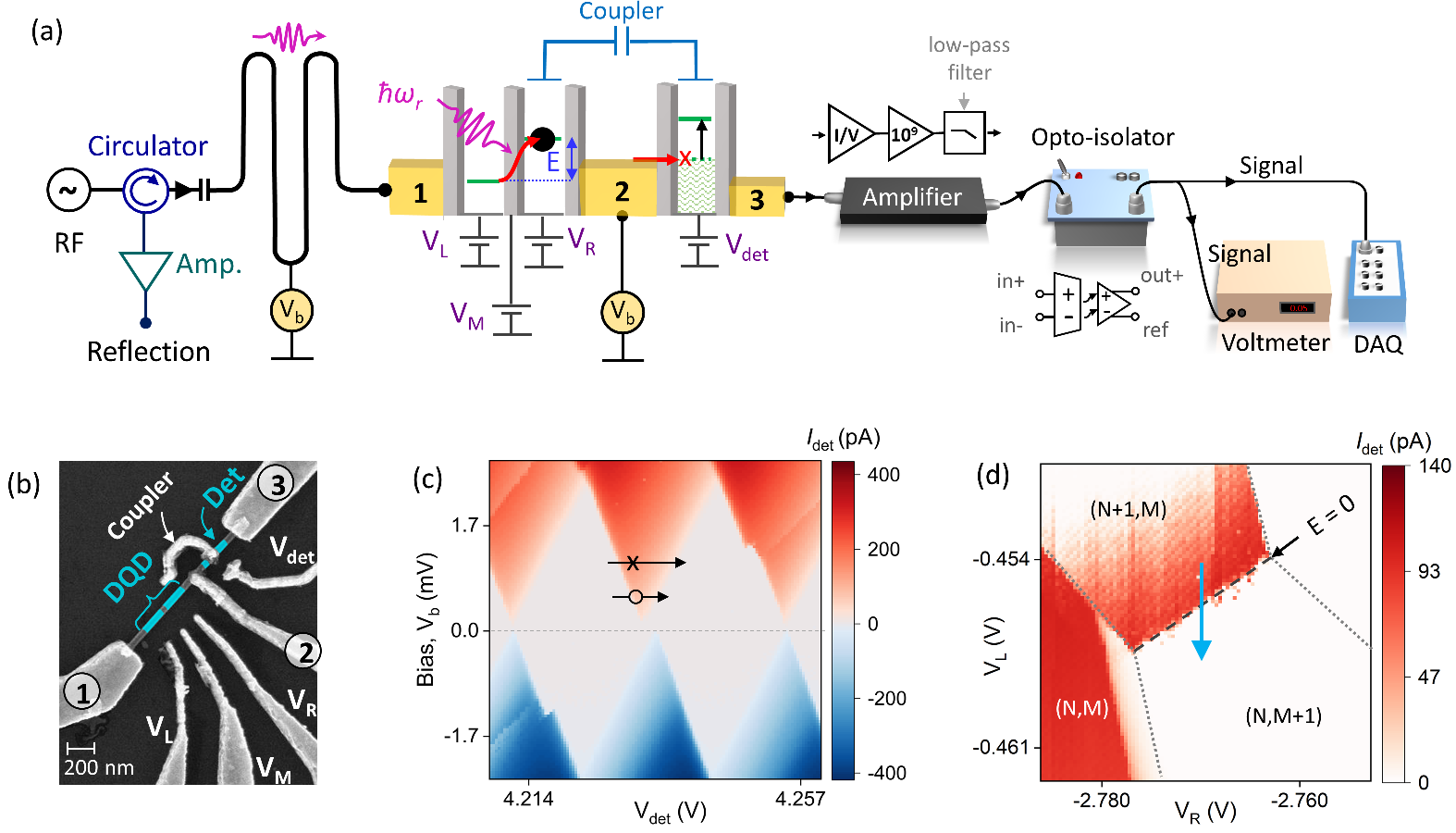}
\caption{\label{fig1}{\bf{Experimental set-up, device operation and DC measurements.}} (a) Schematic representation of the cavity-coupled DQD-QD microwave photon counter. The device utilizes photon-assisted tunneling, where the energy of an incident photon excites an electron from left dot to the right one across the energy level detuning $E$, which is then detected by a third dot capacitively coupled to the DQD. (b) False-colored scanning electron microscope image of the nanowire-QD device, highlighting the DQD and the detector dot. (c) Measured current-bias voltage characteristics of the detector as a function of gate voltage $V_\text{det}$. (d) The charge stability diagram of the DQD showing the detector current $I_\text{det}$ measured as a function of the plunger gate voltages $V_\text{L}$ and $V_\text{R}$. The detector is operated at a charge-sensitive point, marked with a cross in panel c. The $I_\text{det}$ changes abruptly when the electron occupancy in the left or right dot changes. The dotted lines indicate the boundary of constant occupation in the DQD. Moving along the blue arrow in panel d results in an inter-dot charge transfer of (\textit{N}+1, \textit{M}) → (\textit{N}, \textit{M}+1) at level detuning \textit{E} = 0, where \textit{N} and \textit{M} denote the carrier occupancy in left and right QDs.}
\end{figure*}

\begin{figure*}
\includegraphics[width=6.0in]{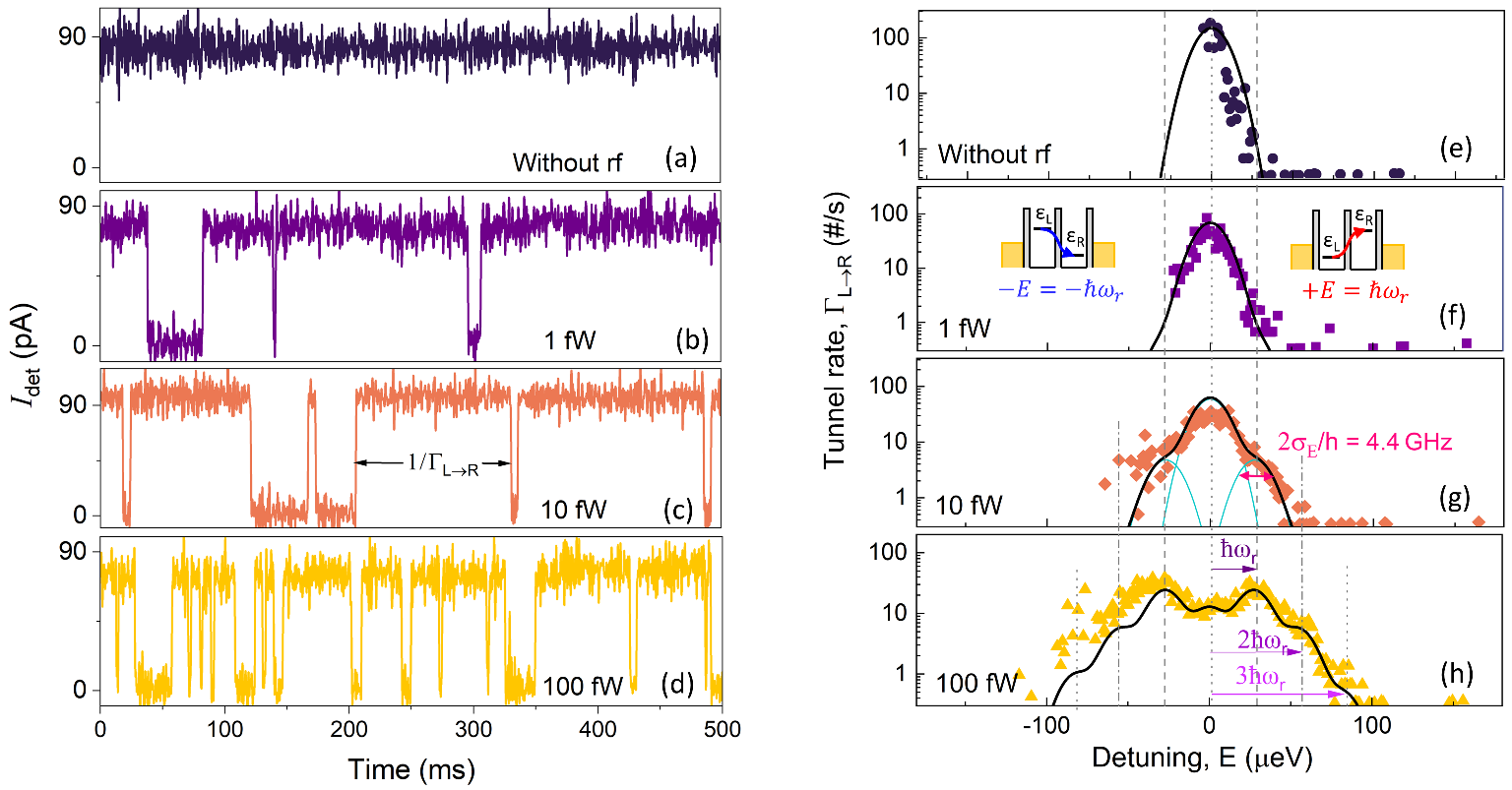}
\caption{\label{fig2}{\bf{Real-time observation of the microwave photon-assisted tunneling of an electron.}} Typical time traces of the detector current $I_\text{det}$ with the DQD level detuning set to $E=\hbar\omega_\text{r}$ = 27.3 $\mu$eV in the absence of the microwave drive (a) and with varying microwave powers (b-d). With microwave power turned on, the $I_\text{det}$ exhibits a two-level telegraph-like signal. When an electron tunnels from QD-L to QD-R across $E=27.3$~$\mu$eV, the $I_\text{det}$ goes to the noise level, and then $I_\text{det}$ jumps up as the electron tunnels back. (e-h) The measured tunnel rate from QD-L to QD-R as a function of the level detuning \textit{E} for the four cases. Insets in panel f illustrate the DQD energy band configurations for the negative and positive detuning cases. The solid black lines depict the theoretical predictions based on the $P(E)$ theory.}
\end{figure*}

Figure~\ref{fig1}(a) schematically illustrates the operation principle of our microwave photon counter. The device utilizes nanowire QDs and a coplanar waveguide resonator. The resonator is directly connected to the DQD and capacitively coupled to a microwave input port. A microwave photon sent via the resonator input causes photon-assisted tunneling of an electron in the DQD absorber, which is energetically permissible when the DQD level detuning, $E=\varepsilon_\text{R}-\varepsilon_\text{L}$, matches the photon energy, see Fig.~\ref{fig1}(a). The energy levels $\varepsilon_\text{R}$ and $\varepsilon_\text{L}$ of the left and right QDs are tuned by the plunger gate voltages $V_\text{L}$ and $V_\text{R}$. The QDs here are formed by periodic switching of the zinc blende and wurtzite phases during nanowire growth~\cite{Luna2015, barker2019, thelander2011, chen2017}. An electron tunneling from the left dot to the right one due to the photon absorption changes the electrostatic potential near the DQD. A third dot (detector), capacitively coupled to the DQD via a metal coupler, see Fig.~\ref{fig1}(b), detects these changes~\cite{barker2022}. We operate the detector at a charge-sensitive point. Recording change in detector current ($I_\text{det}$) thus allows us to probe the photon absorption events in real time.


{\bf{Results.}} We begin by measuring the current-bias voltage Coulomb blockade characteristics of the detector dot as a function of the detector gate voltage $V_\text{det}$. Figure~\ref{fig1}(c) shows the measured results. Here, the bias voltage $V_b$ is applied equally on the leads across the DQD, i.e., lead-1 and 2, and the detector current $I_\text{det}$ is measured via lead-3, located on the other side of the detector. Next, we operate the detector at a charge-sensitive point on the Coulomb peak at $V_b$ = 1.2~mV, marked with a cross in Fig.~\ref{fig1}(c), and map the charge stability diagram of the DQD by measuring $I_\text{det}$ as a function of $V_\text{L}$ and $V_\text{R}$, Fig.~\ref{fig1}(d). Note that larger bias voltage offers here a higher dynamic range of the detector, which is needed for mapping the larger plunger gate space of the DQD in this case. The current changes abruptly when the occupation of the DQD varies, giving rise to a typical honeycomb pattern. The dashed line in Fig.~\ref{fig1}(d) corresponds to zero level-detuning, $E = \varepsilon_\text{R} - \varepsilon_\text{L}$ = 0. The detector current shows a jump of $\Delta I_\text{det}$ = -90 pA when an electron tunnels from QD-L to R, along the blue arrow in Fig.~\ref{fig1}(d), while the total electron occupancy in the DQD remains constant. 

{\bf{Photon counting measurements.}} We now turn to probe the inter-dot tunneling events by continuously measuring the detector current at a charge-sensitive point. We use a lower bias voltage, $V_b$ = 0.6 mV, to minimize the detector back-action effects~\cite{granger2012, li2013}, while maintaining a sufficient signal-to-noise ratio for the photon counting. Using $V_\text{L}$ and $V_\text{R}$, we tune the DQD energy levels with lever arms $\alpha_\text{L}$ = 67~$\mu$eV/mV and $\alpha_\text{R}$ = 21~$\mu$eV/mV. The supplementary information (SI) presents the estimation of lever arms by mapping the finite bias triangles. We measure the time traces as a function of DQD level detuning with different microwave powers. The results are presented in Fig.~\ref{fig2}. Figure~\ref{fig2}(a) shows an exemplary 500~ms time trace recorded at $E=\hbar\omega_\text{r}$ without a microwave drive. Here we observe that the detector current remains constant. We repeat in Figs.~\ref{fig2}(b-d) the measurement with a continuous microwave tone at frequency $\omega_\text{r}$ and varying input power $P_\text{in}$ via the resonator input port. Details of the RF heterodyne circuit and the input power calibration can be found in Refs.~\citealp{khan2021,haldar2023}. In Figs~\ref{fig2}(b-d), $I_\text{det}$ displays a two-level telegraph-like signal. Each time an electron tunnels from QD-L to R, $I_\text{det}$ goes down to the noise level, and goes up by the same amount when the electron relaxes back. We observe the number of jumps in $I_\text{det}$ of Figs.~\ref{fig2}(c-d) increases with higher power.

Figures~\ref{fig2}(e-h) present the tunnel rate $\Gamma_{\text{L}\rightarrow \text{R}}$, counted from 3-second time traces, as a function of detuning $E$ with varying microwave powers. We tune the DQD across the zero-detuning line, as indicated by the blue arrow in Fig.~\ref{fig1}(d). In the absence of microwave drive in Fig.~\ref{fig2}(e), $\Gamma_{\text{L}\rightarrow \text{R}}$ peaks around \textit{E} = 0 when $\varepsilon_\text{L}$ and $\varepsilon_\text{R}$ are aligned~\cite{hofmann2020}. As we turn on the microwave drive, a new shoulder forms on both sides of the resonance peak at $P_\text{in}$ = 10~fW in Fig.~\ref{fig2}(g), which eventually evolves into two distinct peaks at 100~fW in Fig.~\ref{fig2}(h). Furthermore, in Fig.~\ref{fig2}(h), we observe another set of shoulders appearing at higher detuning. We find that the spacing of these shoulders on either side of the zero-detuning corresponds to the single-photon energy, $\Delta E=\hbar\omega_\text{r}$. Therefore, the photon energy sets the relevant energy scale of the measured tunnel rate.

\begin{figure}
\includegraphics[width=3.2in]{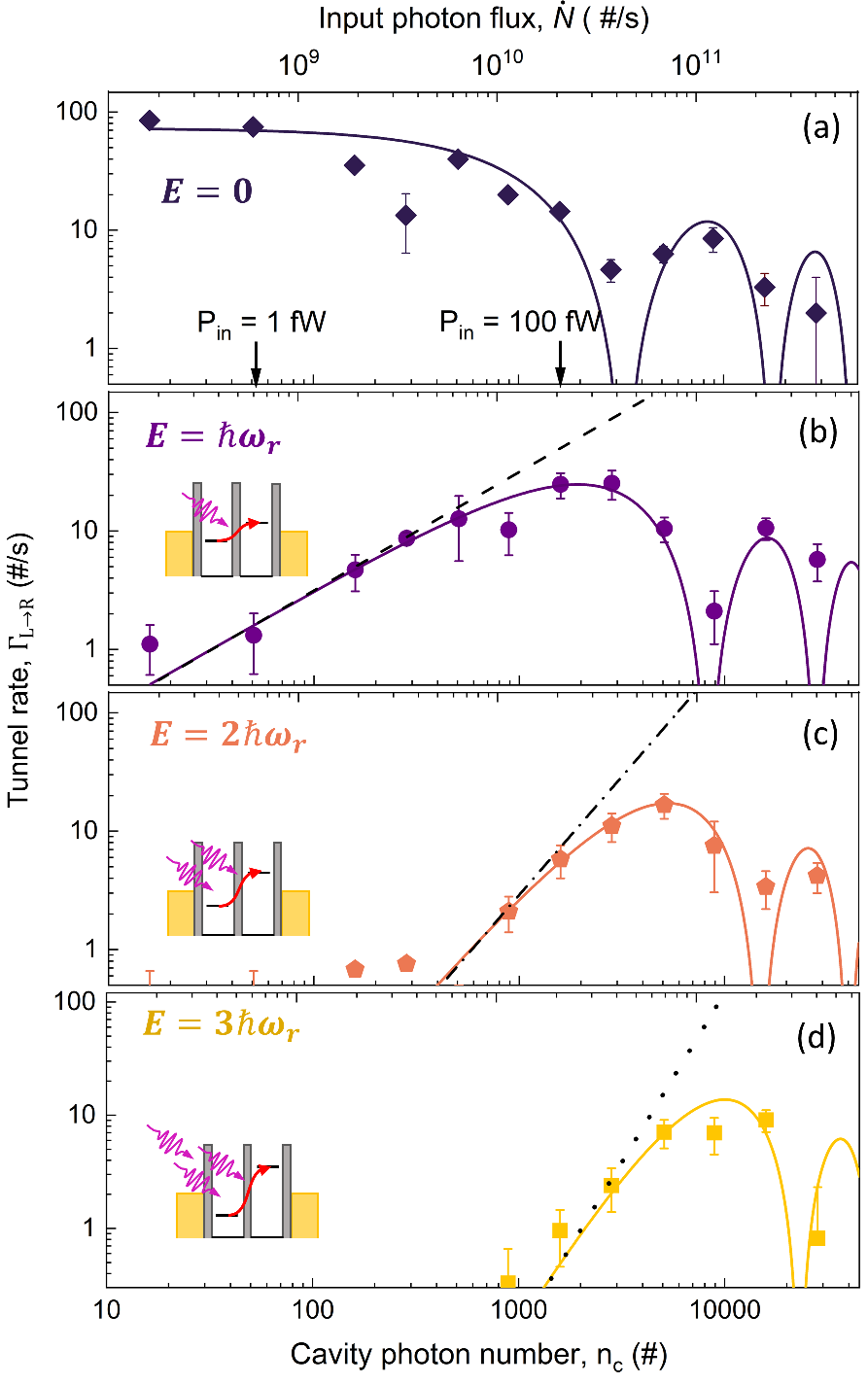}
\caption{\label{fig3}{\bf{Tunnel rates modulated by the number of photons stored in the cavity.}} (a-d) The measured tunnel rate $\Gamma_{\text{L}\rightarrow\text{R}}$ as a function of the cavity photon number $n_\text{c}$ for different DQD level detuning: \textit{E} = 0, $\hbar\omega_\text{r}$, $2\hbar\omega_\text{r}$, and $3\hbar\omega_\text{r}$ respectively.  The scale bar above indicates the corresponding photon flux $\Dot{N}$ at the input port of the resonator. The solid lines represent theoretical predictions based on the $P(E)$ theory. The black dashed lines show the $\Gamma_{\text{L}\rightarrow \text{R}} \propto n_\text{c}^k$ dependency where $k$ = 1, 2 and 3 respectively.}
\end{figure}

Figures~\ref{fig3}(a-d) present the measured $\Gamma_{\text{L}\rightarrow\text{R}}$ at $E=k\cdot\hbar\omega_\text{r}$, with k = 0 to k = 3, as a function of $P_\text{in}$. We also indicate the corresponding cavity photon number, $n_\text{c}=2C_\text{r}Q_\text{L}^2Z_0P_\text{in}/Q_\text{ex}\hbar\omega_\text{r}$~\cite{haldar2022,havir2023} and photon flux at the resonator input, $\dot{N}=P_\text{in}/\hbar\omega_\text{r}$, in Fig.~\ref{fig3}. Here, $C_\text{r}$ = 540~fF is the total capacitance of the resonator, $Q_\text{ex}$ = 3350 the external quality factor, $Q_\text{L}$ = 2050 the loaded quality factor, and $Z_0$ = 58~$\Omega$ represents the impedance of the resonator. In Fig.~\ref{fig3}(a), the inter-dot tunnel rate measured with $E=0$ shows more than an order of magnitude suppression with higher input powers. Nonetheless, the tunnel rate measured at $E=\hbar\omega_\text{r}$ in Fig.~\ref{fig3}(b) increases linearly with the cavity photon number $n_\text{c}$ and then exhibits an oscillatory decrease under a strong microwave drive. The dashed black line indicates this linear dependency. With detuning $E=2\cdot\hbar\omega_\text{r}$ in Fig.~\ref{fig3}(c), tunneling events become measurable when $n_\text{c}$ exceeds $\sim$200. Here, $\Gamma_{\text{L}\rightarrow\text{R}}$ increases quadratically with $n_\text{c}$, as shown by the dashed-dotted line. Following the same trend, at $E=3\cdot\hbar\omega_\text{r}$ in Fig.~\ref{fig3}(d), $\Gamma_{\text{L}\rightarrow\text{R}}$ becomes measurable at $n_\text{c}>1000$ with a cubic dependence on $n_\text{c}$, as indicated by the dotted line. In all four cases, $\Gamma_{\text{L}\rightarrow\text{R}}$ shows overall suppression under strong microwave drive.

In the absence of microwave drive in Fig.~\ref{fig2}(e), the tunneling rate is low at $E = k\cdot\hbar\omega_\text{r}$ when $k\neq0$. This is because the electronic temperature of 60~mK~\cite{barker2019} corresponds to much smaller energy scale $k_\text{B}T_\text{e}$ = 5.2~$\mu$eV than the photon energy of $\hbar \omega_\text{r}$ = 27.3~$\mu$eV. It is therefore evident that the measured $\Gamma_{\text{L}\rightarrow\text{R}}$ in Figs.~\ref{fig3}(b-d) primarily originate from the photon-assisted tunneling in the DQD. The signatures of photon-assisted tunneling  on the stability diagram, previously reported in Refs.~\citealp{petta2004, gustavsson2007}, are also observed in our measurements, see the SI.

{\bf{The $P(E)$ theory.}} Using Fermi's golden rule~\cite{ingold1992, souquet2014, tien1963, devoret1990}, the general expression of the photon-assisted tunnel rate reads:

\begin{equation}\label{rate}
\Gamma_{\text{L}\rightarrow\text{R}}(E)=\frac{2\pi}{\hbar}t_{\text{L,R}}^2P(E).
\end{equation}

Here, $t_\text{L,R}$ is the tunnel coupling between $\varepsilon_\text{L}$ and $\varepsilon_\text{R}$ of the DQD~\cite{hofmann2020}. Function $P(E)$ describes the probability of absorbing energy $E$ by the environment during the inter-dot tunneling with the level detuning \textit{E}. For an ideal single mode resonator, we have $P(E)=\sum_k J_k^2(2\sqrt\rho\alpha_0)\delta(E-k\hbar\omega_\text{r})$~\cite{souquet2014, tien1963}. This yields the photon-assisted tunnel rate~\cite{roulleau2011, silveri2017, ingold1992}:

\begin{equation}\label{eq2}
   \Gamma_{\text{L}\rightarrow\text{R}}(E) =  \frac{2\pi}{\hbar} t_\text{L,R}^2 \sum_{k=0}^{\infty} J_k^2(2\sqrt\rho_\text{cav} \alpha_0) \delta(E-k\hbar\omega_\text{r}), 
\end{equation}

where, $J_k(2\sqrt\rho_{cav} \alpha_0)$ is the k-th order Bessel function, derived from the phase-phase correlation function for the coherent microwave drive with amplitude $\alpha_0=\sqrt{n_\text{c}}$~\cite{souquet2014, gerry2005}. The function is also known as the Tien-Gordon coefficient~\cite{tien1963}, which describes the probability for an absorption process involving $k$ photons. The $\rho_\text{cav}= \xi^2\times\pi Z_0/R_Q$ is the strength of the zero-point voltage fluctuation of the cavity-mode~\cite{silveri2017}, with $R_Q=h/e^2$ the resistance quantum. The parameter $\xi$ defines the coupling strength between the cavity photon modes and the dipole moment of the DQD~\cite{childress2004, silveri2017}.

Next, we fit the measured data of Fig.~\ref{fig3}(b) using equation~(\ref{eq2}) with $t_\text{L,R}$ and $\xi$ as the fitting parameters. Here, $t_\text{L,R}$ determines the overall scale for $\Gamma_{\text{L}\rightarrow\text{R}}$, and $\xi$ sets the overall range of the linear response and the period of the high-power oscillations. In addition, we used here Gaussian broadening of the delta functions with variance $\sigma_\text{E}/h$ = 2.2~GHz, based on the width of the features in Fig.~\ref{fig2}(g) (see the green curves). The broadening $\sigma_\text{E}$ primarily arises due to the decoherence of the DQD, where the decoherence rate $\Tilde{\Gamma}$ = $\sigma_\text{E}/h$ = $2\pi\:\times\:$350~MHz~\cite{petta2004}. This value matches the previously determined decoherence rates of similar DQDs~\cite{khan2021, ranni2023}. The solid line in Fig.~\ref{fig3}(b) shows the fitted result, which yield $t_\text{L,R}$ = 0.43~neV and $\xi$ = 0.26. The fitted value of $\xi$ suggests that the microwave cavity field is divided roughly equally among the three tunnel junctions of the DQD~\cite{haldar2022}. The slight deviation from the value 1/3 could be attributed to gate capacitances lowering the overall coupling strength. In Figs.~\ref{fig3}(a) and \ref{fig3}(c-d), the solid curves show the predictions of the $P(E)$ theory based on the fitted values of $t_\text{L,R}$ and $\xi$. The solid black lines in Figs.~\ref{fig2}(e-h) also depict the predicted $\Gamma_{\text{L}\rightarrow\text{R}}$ as a function of level detuning $E$, obtained using equation~(\ref{eq2}). The theory predictions are found to be in excellent agreement with the experimental data.

At the low-power limit, the photon assisted tunneling rate  can be described by Taylor expanding equation~(\ref{eq2}) for $k = 1$, where the first-order term delineating the linear response regime reads: $\Gamma_{\text{L}\rightarrow\text{R}} \propto t_\text{L,R}^2\rho_\text{cav} n_\text{c}/h\sigma_\text{E}$. The dashed black line in Fig.~\ref{fig3}(b) represents the $P(E)$ theory result with this lowest order term. The above relation of $\Gamma_{\text{L}\rightarrow\text{R}}$ yields the single photon counting efficiency (see the SI), as

\begin{equation}\label{pe3}
\eta = \frac{\Gamma_{\text{L}\rightarrow\text{R}}}{\dot{N}}=\left(\frac{4g^2}{\sigma_\text{E}/h}\right)\frac{4\kappa_c}{\kappa^2},
\end{equation}

where, $g = g_0\cdot 2t_\text{L,R}/\hbar\omega_\text{r}$ is the cavity-DQD coupling constant with the bare coupling constant $g_0 \sim \omega_\text{r}\xi\sqrt{2Z_0/R_Q}$~\cite{childress2004}, input coupling $\kappa_c=\omega_\text{r}/Q_\text{ex}$ and the total
cavity photon decay rate $\kappa=\omega_\text{r}/Q_L$. Equation~(\ref{pe3}) shows that the single photon detection efficiency quadratically increases with the coupling constant, $g$ (or, the tunnel coupling). Thus, a plausible approach to enhance the counting efficiency would be to operate the device in a higher tunnel coupling regime, or by increasing the cavity-DQD coupling strength using high-impedance resonators~\cite{scarlino2021,ranni2023}. We discuss this further in connection with the Jaynes-Cummings theory results, see below. 

{\bf{Connection between the $P(E)$ theory and Jaynes-Cummings input-output theory.}} Here, we compare the above $P(E)$ theory results to the Jaynes-Cummings input-output theory which is often used to describe cavity-DQD interactions~\cite{childress2004, zenelaj2022, liu2014, haldar2023, wong2017, khan2021}. We use the input-output theory model presented in Refs.~\citealp{wong2017, khan2021}. Our photocounter has two important advantages over the earlier photodiode devices~\cite{khan2021}. Firstly, our device can measure both the photon-assisted tunnel rate and its subsequent relaxation rate separately. Therefore, the relaxation does not impact the photon-counting efficiency, provided that the detector bandwidth is sufficiently high to detect each event individually. Secondly, electron tunneling into and out of the DQD is suppressed across the charge transfer line, and thus the directivity (see Ref.~\citealp{khan2021}) of our photodiode does not lower the detection efficiency. Under these conditions, the photon-assisted tunnel rate at low input photon flux is

\begin{equation}\label{eq3}
\Gamma_{\text{L}\rightarrow\text{R}}(E=\hbar\omega_\text{r})=\dot{N}\frac{\kappa_c}{\kappa}\frac{4\kappa_\text{DQD}\kappa}{(\kappa_\text{DQD}+\kappa)^2}.
\end{equation}

We use equation~(\ref{eq3}) to fit $\Gamma_{\text{L}\rightarrow\text{R}}$ as a function of $\dot{N}$ in the linear response regime of Fig.~\ref{fig3}(b) (dashed black line), and calculate the photon dissipation rate due to the DQD absorption $\kappa_\text{DQD} = 4g^2/\Tilde{\Gamma}$ = $2\pi\:\times\:$$3.5\cdot10^{-3}$ Hz. Here we used $\kappa_c$ = $2\pi\:\times\:$2.0~MHz and $\kappa$ = $2\pi\:\times\:$3.2~MHz, obtained using RF reflectivity measurements, see the SI. With the above values of $t_\text{L,M}$ and $\Tilde{\Gamma}$ obtained with the $P(E)$ theory and the above value of $\kappa_\text{DQD}$ obtained using input-output theory, we estimate the cavity-DQD coupling constant, $g=(\kappa_\text{DQD}\Tilde{\Gamma}/4)^{1/2}$ = $2\pi\:\times\:$0.6~kHz and bare coupling constant $g_0$ = $2\pi\:\times\:$24~MHz. This is in excellent agreement with the previously reported coupling constant for a similar coplanar waveguide resonator-coupled DQD~\cite{khan2021, frey2012, liu2014}. As an additional support, we performed RF spectroscopy on our device in a higher tunnel coupling regime (see, the SI) and obtained $g_0$ = $2\pi\:\times\:$45~MHz. These findings imply that under the operation conditions of our device, the $P(E)$ theory and input-output formalism are consistent with each other.

Next, using the linear response regime in Fig. \ref{fig2}(b), we obtain single photon counting efficiency, $\eta$ = $\Gamma_{\text{L}\rightarrow\text{R}} / \dot{N}$ = $10^{-7}$\:\%. Using equation~(\ref{eq3}), the quantum efficiency of single photon counting reads: $\eta = 4\kappa_c\kappa_\text{DQD}/(\kappa_\text{DQD} + \kappa)^2$. With $\kappa_\text{DQD} = 4g^2/\Tilde{\Gamma}$ and small internal losses in the resonator, the above theoretical relation of $\eta$, derived from the input-output theory, shows excellent consistency with the results obtained from the $P(E)$ theory, as seen in equation~(\ref{pe3}). The extremely low quantum efficiency arises from the nine orders of magnitude smaller $\kappa_\text{DQD}$ when compared to $\kappa_c$ or $\kappa$. This is a direct consequence of the sub-nano-eV tunnel coupling, which has led to a small cavity-DQD coupling, $g$ = $2\pi\:\times\:$0.6~kHz, and a correspondingly diminutive $\kappa_\text{DQD}$. Therefore, a promising approach to achieve near unity quantum efficiency would be to employ a fast detection scheme instead of the DC measurement, enabling the device operation with an increased tunnel coupling. With a tunnel coupling of $t_{\text{L,R}}$ = 4~$\mu$eV, while keeping all other parameters constant and with input power $P_\text{in}$ = 1~fW, the photon-assisted tunnel rate increases to $\sim$100~MHz, resulting in $\eta$ = 50\:\%. However, charge detection with approximately 100~MHz bandwidth requires further investigation, potentially through the use of an RF-based measurement approach~\cite{Liu2021, Connors2020, havir2023, Florian2023}.

In summary, we have presented a continuous microwave photon counter capable of detecting both single and multiple photon absorption events in photon-assisted tunneling. We measured photon absorption rates with varying cavity photon numbers and analyzed the results in the environmental $P(E)$ theory framework. The input microwave photons set a tuneable environment for the cavity photon field, which we were able to measure using our photon counter. Our findings show that the pioneering Tien-Gordon theory~\cite{tien1963}, originally developed for the electrons tunneling under a purely classical AC drive, remains applicable in the quantum regime with low power levels, where the particle nature of microwaves is expected to dominate over their wave character~\cite{haldar2022}. Our study provides a foundation for treating the environmental $P(E)$ theory predictions as quantitatively equivalent to the Jaynes-Cummings input-output theory results at the single photon detection limit. With a fast readout scheme, our results lay the avenue for continuous high-efficiency photocounting in the microwave domain, holding significant promise for applications in quantum information processing, quantum communication, and quantum optics. 

\subsection*{Methods}

The device was fabricated on a high-resistivity silicon substrate ($>$10~k$\Omega$-cm) with a 200~nm thick thermally grown SiO$_2$ layer. The resonator, along with the DC lines, were fabricated using a 100~nm thick sputtered Nb film. The co-planar waveguide resonator has a fundamental resonance at $\omega_\text{r}$ = $2\pi\:\times\:$6.615~GHz. To prevent microwave photon leakage through the gate lines, we applied a 30~nm Al$_2$O$_3$/50~nm Al ground plane directly on the DC gate lines. This capacitive coupling of the ground plane with the DC lines effectively reduces the impedance of the gate lines to approximately 50~m$\Omega$, which is three orders of magnitude smaller than the resonator impedance. The Al$_2$O$_3$ layer was deposited using the atomic layer deposition technique. Photolithography and the lift-off method were used for both resonator and ground plane fabrication. Next, the nanowires, grown by metal-organic vapour phase epitaxy, were transferred from the growth chip to the device chip. The positions of the randomly transferred nanowire quantum dots were determined using scanning electron microscopy imaging. Finally, source-drain contacts and gate lines were made by electron beam lithography, followed by the evaporation of 30~nm Ni/120~nm Au films. The Ti/Au pads were used for interconnections between Nb lines and Ni/Au lines connecting the nanowire device. Additional details about the device fabrication process can be found in Ref.~\citealp{barker2019}. 

The measurements were performed in a dilution refrigerator at an electronic temperature of 60~mK. Using the plunger gate voltages $V_\text{L}$, $V_\text{R}$ and $V_\text{M}$, we decrease the tunnel rates to less than 1~kHz, which enables the detection of individual tunnel events by measuring the electric current using a nearby charge detector. For the presented experiments, we set $V_\text{M}$ = -2.7~V unless otherwise specified. For the DC measurements, the detector current was amplified ($\times10^9$ V/A) using a DLPCA-200 amplifier with a 1~kHz bandwidth, see Fig.~\ref{fig1}(a). The amplified signal then passed through an opto-electronic isolator and then signal was measured using Keysight 34461A voltage meter. For time trace measurements, the same signal after the opto-electronic isolater was recorded using a data acquisition (DAQ) card. We digitally filtered the data with a 0.8~kHz low-pass filter to limit noise contribution in our analysis. The digitized trace was acquired by defining a threshold level ($I_\text{T}$) of the detector current as the mid-point between the two main current states. When $I_\text{det} < I_\text{T}$, the electron was in the right dot (photon absorbed); otherwise, it was in the left dot (relaxation state). Further details on data analysis can be found in the supplementary information of Refs.~\citealp{ranni2021, barker2022}.

\bibliography{master}

\subsection*{Acknowledgments}
We acknowledge fruitful discussions with Claes Thelander, Peter Samuelsson, Patrick Potts, Drilon Zenelaj, Adam Burke, and Martin Leijnse and the financial support from the Knut and Alice Wallenberg Foundation through the Wallenberg Center for Quantum Technology (WACQT), the European Union (ERC, QPHOTON, 101087343), NanoLund, and Swedish Research Council (Dnr 2019-04111). Views and opinions expressed are however those of the author(s) only and do not necessarily reflect those of the European Union or the European Research Council Executive Agency. Neither the European Union nor the granting authority can be held responsible for them. We also thank Jonas Bylander, Simone Gasparinetti, and Anita Fadavi Roudsari from the Chalmers University of Technology for fruitful discussion to improve the resonator and providing high quality factor reference resonator devices.

\subsection*{Author Contributions}
V.F.M. conceived the experiment. S.L. and K.D.T. designed and fabricated the nanowires. V.F.M. and S.H. designed and fabricated the device. S.H. performed the measurements. S.H. and V.F.M. performed the data analysis. All authors contributed to the discussion of the results and the manuscript preparation.

\subsection*{Competing Interests}
The authors declare no competing interests.

\subsection*{Data availability}
The data that support the findings of this study are available from the authors upon reasonable request.

\end{document}


\title{Supplementary Information: Continuous microwave photon counting by semiconductor-superconductor hybrids}

\author{Subhomoy Haldar}
\email{subhomoy.haldar@ftf.lth.se}
 \address{NanoLund and Solid State Physics, Lund University, Box 118, 22100 Lund, Sweden}
   \author{David Barker}
 \address{NanoLund and Solid State Physics, Lund University, Box 118, 22100 Lund, Sweden}
  \author{Harald Havir}
 \address{NanoLund and Solid State Physics, Lund University, Box 118, 22100 Lund, Sweden}
   \author{Antti Ranni}
 \address{NanoLund and Solid State Physics, Lund University, Box 118, 22100 Lund, Sweden}
 \author{Sebastian Lehmann}
 \address{NanoLund and Solid State Physics, Lund University, Box 118, 22100 Lund, Sweden}
\author{Kimberly A. Dick }
 \address{NanoLund and Solid State Physics, Lund University, Box 118, 22100 Lund, Sweden}
 \address{Center for Analysis and Synthesis, Lund University, Box 124, 22100 Lund, Sweden}
 \author{Ville F. Maisi}
  \email{ville.maisi@ftf.lth.se}
 \address{NanoLund and Solid State Physics, Lund University, Box 118, 22100 Lund, Sweden}
 
\date{\today}

\maketitle

Further information is provided here on the calculation of the DQD lever arms and cavity photon dissipation rates of the device. We present charge stability maps that exhibit the signature of photon-assisted tunneling on both sides of the zero-detuning line. Furthermore, to determine the bare-cavity-DQD coupling constant of the DQD, we increased the tunnel coupling of the DQD and performed RF spectroscopy measurements. The results of DC transport measurements on the DQD together with RF spectroscopy are presented. Finally, the $P(E)$ theory results in the linear response regime of the single photon detection are discussed in connection with the Jaynes-Cummings input-output theory.

\subsection{Mapping finite bias triangles and estimation of the DQD lever arms.}

\begin{figure}[h]
\includegraphics[width=5.3in]{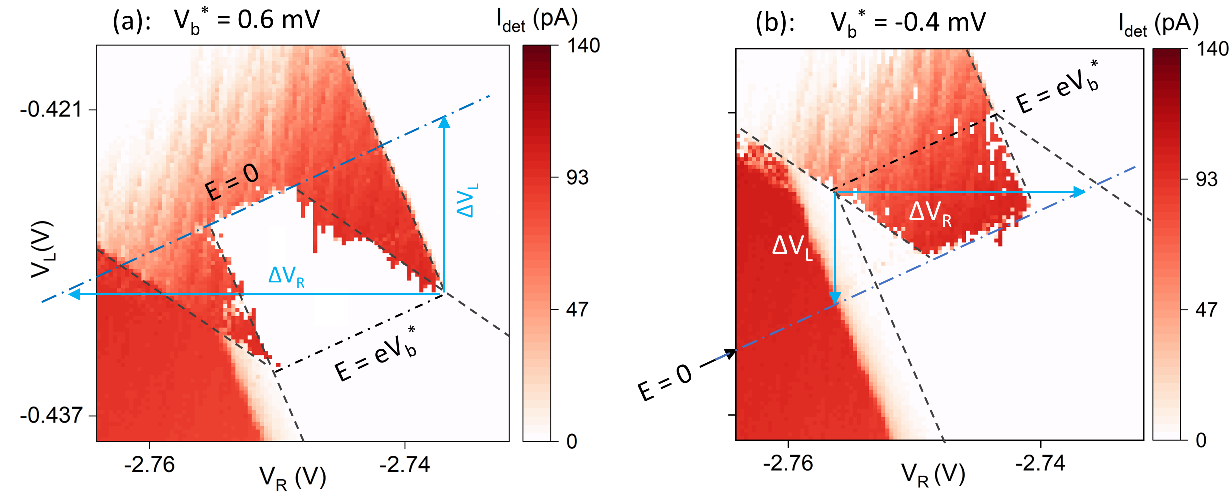}
\caption{\label{figS1}\textbf{Finite bias triangles and estimation of the DQD lever arms.} The charge stability diagram of the DQD showing the detector current $I_\text{det}$ measured as a function of the plunger gate voltages $V_\text{L}$ and $V_\text{R}$ with the bias voltage $V_b^*$ = 0.6~mV (a) and $V_b^*$ = -0.4~mV (b) applied across the DQD. The dashed black lines indicate the boundary of the constant occupation region in the stability map and blue dash-dotted line indicate the base of the FBTs with detuning $E=0$.}
\end{figure}

 We measure charge stability diagram of the DQD by recording the detector current $I_\text{det}$ as a function of $V_\text{L}$ and $V_\text{R}$ while applying a bias voltage $V_b^*$ across the DQD. Once agian, the detector is operated at a charge-sensitive point on the Coulomb peak at $V_b$ = 1.2~mV, marked a cross in Fig.~1(c) of the main text. In this configuration, the bias voltage $V_b^*$ across the DQD is determined as the difference between the voltages applied to lead 2 ($V_{b,2}$) and lead 1 ($V_{b,1}$) of the device, i.e., $V_b^* = V_{b,2} - V_{b,1}$, whereas voltage applied to lead 3 ($V_{b,3}$) sets the bias voltage applied to the detector $V_b = V_{b,3} - V_{b,2}$. Figure~\ref{figS1} shows the measured result with $V_b^*$ = 0.6~mV and $V_b^*$ = -0.4~mV. The triangular areas enclosed by the dashed black lines represent the finite bias triangles (FBTs), a characteristic transport feature of the DQDs~\cite{shin2010, barker2019, ihn2009}. The blue dash-dotted line indicates the base of these triangles when the level detuning is zero, i.e., $E=\varepsilon_\text{R}-\varepsilon_\text{L}=0$. With a bias voltage $V_b^*$ applied across the DQD, an energy window of $eV_b^*$ is opened for the charge transport. The tip of the FBTs corresponds to the situation when the level detuning matches the energy bias window~\cite{khan2021}. Therefore, the black dash-dotted line, passing through the tip of the FBTs, corresponds to $E= eV_b^*$ = 0.6~meV in Fig.~\ref{figS1}(a) and $eV_b^*$ = -0.4~mV in Fig.~\ref{figS1}(b). Since the plunger gate voltages linearly move the energy levels of the respective dots, measuring the size of the FBTs as shown in Fig.~\ref{figS1}, yields the lever arms along the detuning direction: $\alpha_\text{L} = V_b^*/\Delta V_\text{L}$ = 66.6~$\mu$eV/mV and $\alpha_\text{R} = V_b^*/\Delta V_\text{R}$ = 20.5~$\mu$eV/mV~\cite{vanderWiel2002, taubert2011}.
 
\subsection{Signatures of photon assisted tunneling on the DQD charge stability map.}\label{AppenPhotonMap}

\begin{figure}[h]
\includegraphics[width=6.0in]{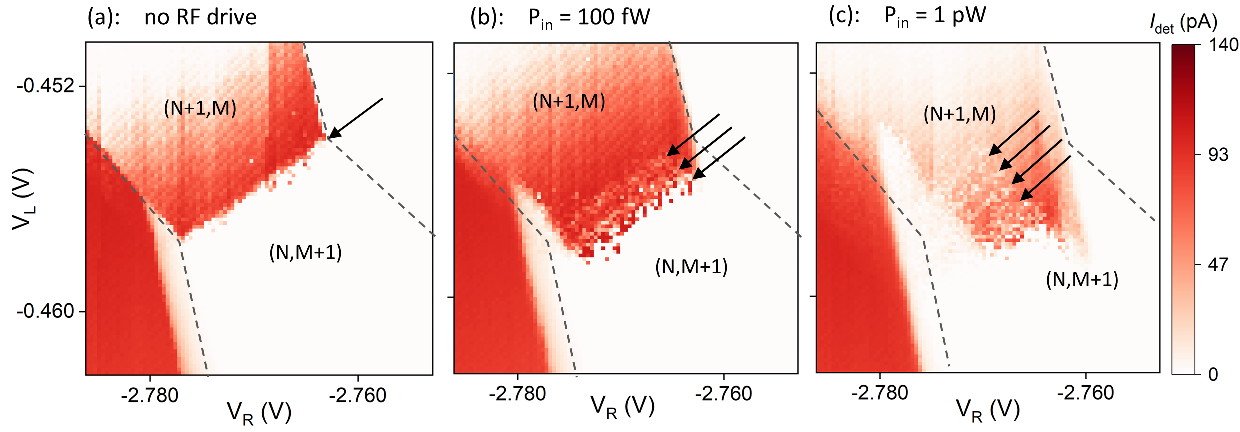}
\caption{\label{figS2}\textbf{The signatures of photon assisted tunneling on the charge stability map.} Charge stability diagrams of the DQD showing the detector current $I_\text{det}$ as a function of plunger gate voltages $V_\text{L}$ and $V_\text{R}$ without the microwave drive (a), and with input power $P_\text{in}$ = 100~fW (b) and 1~pW (c). Dashed lines indicate the boundaries of the constant occupation region in the DQD. Strong microwave drive results in the appearance of multiple lines near the interdot charge transfer line, indicated by black arrows. }
\end{figure}

We measure the charge stability map of the DQD using the detector dot under identical settings as the results presented in Fig.~1(d). We do not apply a bias voltage across the DQD. Figure~\ref{figS2}(a) shows the measured result without a microwave drive applied to the device. With input power $P_\text{in}$ = 100~fW in Fig.~\ref{figS2}(b), we observe that multiple lines close to the zero-detuning line appear. When we increase the input power to 1~pW in Fig.~\ref{figS2}(c), the number of lines visible at the inter-dot charge transfer line increases. We note that the separation of these lines roughly corresponds to the photon energy of the drive, suggesting the effects of photon assisted tunneling in the DQD. The multiple lines appearing close to the zero-detuning line, as a signature of photon-assisted tunneling, have been previously reported when probing the charge stability diagram of DQDs using a quantum point contact~\cite{petta2004}. The corresponding oscillations (or shoulders) in the tunnel rate across the inter-dot charge transfer line is presented in Fig.~2 of the main text.

\subsection{RF spectroscopy and estimation of cavity photon dissipation rates.}\label{AppenKappaEstimation}
\begin{figure}[b]
\includegraphics[width=2.5in]{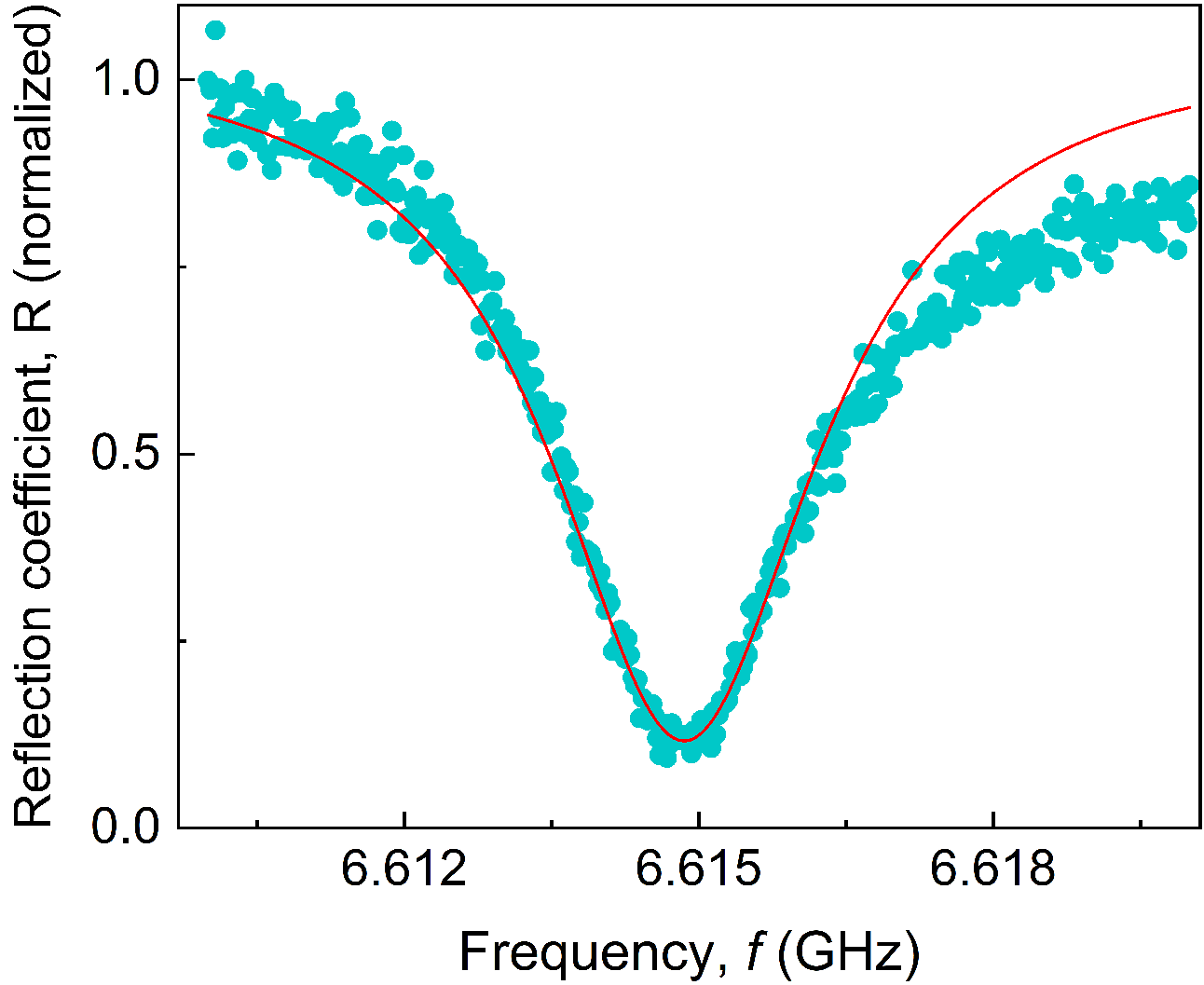}
\caption{\label{figS3}\textbf{RF reflectance spectra.} The measured reflection coefficients as a function of drive frequency at the Coulomb blockade regime of the DQD. The solid red curve shows the theoretically fitted result to estimate cavity photon dissipation rates.}
\end{figure}
Figure~\ref{figS3} shows the measured reflection coefficient, \textit{R}, as a function of the drive frequency, \textit{f}. To avoid additional contribution due to the DQD absorption, we tune the DQD to a Coulomb blockade regime. The measurement is performed with an input power of $P_\text{in}$ = 1~fW. To obtain the input coupling $\kappa_c$ and internal losses $\kappa_\text{in}$ of our one-port resonator, we fit the measured data using~\cite{khan2021}:

\begin{equation}
    R=\frac{(\kappa_\text{in}-\kappa_c)^2+4(\omega-\omega_\text{r})^2}{(\kappa_\text{in}+\kappa_c)^2+4(\omega-\omega_\text{r})^2},
\end{equation}

where $\omega_\text{r}=2\pi f_r$ represents the resonance frequency of the cavity. The solid red curve in Fig.~\ref{figS3} shows the fitted result, which yield $\kappa_c$ = $2\pi\:\times\:$2.0~MHz and $\kappa_\text{in}$ = $2\pi\:\times\:$1.2~MHz. Therefore, the total cavity photon dissipation rate of our device is, $\kappa=(\kappa_c+\kappa_\text{in})$ = $2\pi\:\times\:$3.2~MHz.

\subsection{Estimation of cavity-DQD coupling constant at higher tunnel coupling regime.}\label{AppenCouplingConst}

\begin{figure*}[b]
\includegraphics[width=6.0in]{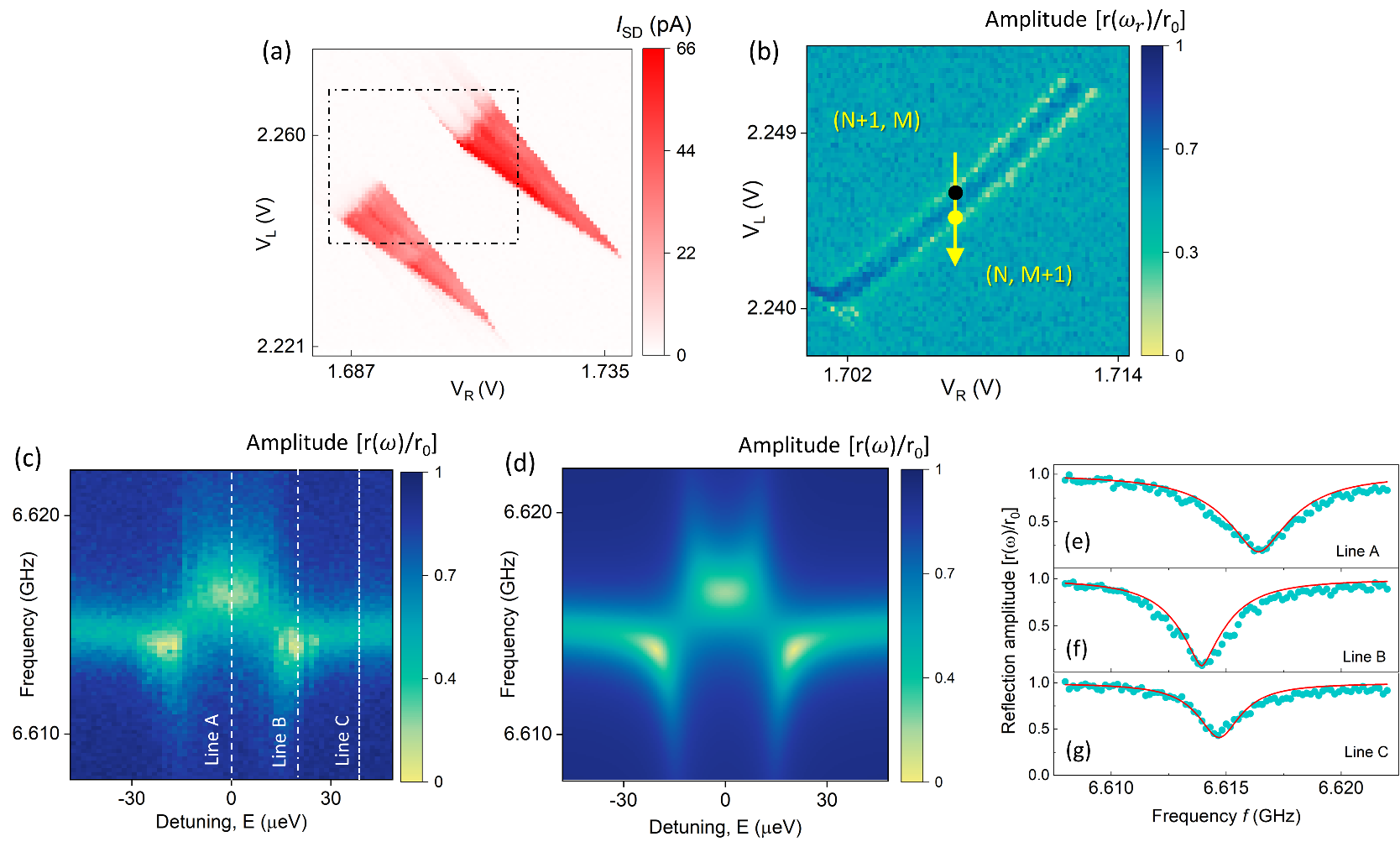}
\caption{\label{figS4}\textbf{The results obtained by DC and RF measurements at higher tunnel coupling regime.} (a) Measured current running through the DQD as a function of plunger gate voltages with bias voltage $V_b$ = 1~mV and (b) normalized RF amplitude $r(\omega_\text{r})/r_0$ as a function of plunger gate voltages $V_\text{L}$ and $V_\text{R}$, with $V_\text{M}$ = 1~V. (c) Measured RF amplitude as a function of drive frequency \textit{f} and detuning \textit{E}, along the arrow indicated in panel c. The measurements are performed with input power, $P_\text{in}$ = 1 fW. (d) Theoretically fitted response using equations~\ref{eqS1} and \ref{eqS2}. (e), (f) and (g) show the experimental data corresponding to the line-cuts along line A, B and C as indicated in panel (c). Red solid curves show the corresponding theoretically fitted results. }
\end{figure*}

The $\kappa_\text{DQD}$, obtained in the gate-voltage regime discussed in the main text, is significantly smaller than $\kappa_c$ and $\kappa_\text{in}$. As a result, the effects of DQD absorption on the RF reflectance becomes vanishingly small. Consequently, it is not possible to determine cavity-DQD coupling constant with an exceedingly small tunnel coupling $t_\text{L,R}$ using RF spectroscopy. 

Therefore, we tune the DQD into a higher tunnel coupling regime by adjusting three plunger gate voltages $V_\text{L}$, $V_\text{M}$, and $V_\text{R}$. We here used $V_\text{M}$ = 1~V. This also allows us to measure the FBTs directly by recording the current $I_\text{SD}$ running through the DQD with an applied source-drain bias voltage ($V_b^*$). Figure~\ref{figS4}(a) shows the measured result with $V_b^*$ = 1~mV, which we use to calculate the lever arms $\alpha_\text{L}$ = 20.5~$\mu$eV/mV and $\alpha_\text{R}$ = 18.5~$\mu$eV/mV. 
At the resonance condition, when the photon energy $\hbar \omega_\text{r}$ matches the energy gap between the discrete energy levels of the DQD $(E^2+4t_\text{L,R}^2)^{1/2}$, the DQD introduces a loss channel to the cavity-photon mode. Additionally, the coherent interaction between the DQD levels and the microwave photons causes a dispersive effect. To investigate these effects, first, we drive the system with $\omega = \omega_\text{r}$ = $2\pi\:\times\:$6.615~GHz and measure the reflection amplitude as a function of the gate voltages $V_\text{L}$ and $V_\text{R}$. In Fig.~\ref{figS4}(b), the measured results are shown with input power $P_\text{in}$ = 1~fW and bias voltage $V_b^*$ = 0. This is measured within the plunger gate voltage range enclosed by the dash-dotted line in Fig.~\ref{figS4}(a). We observe two distinct absorption lines in Fig.~\ref{figS4}(b), corresponding to the resonant tunneling of (\textit{M}, \textit{N}+1) $\Leftrightarrow$ (\textit{M}+1, \textit{N}) at $E_r=\pm\hbar\omega_\text{r}$ (black and yellow dots).

Next, in Fig.~\ref{figS4}(c), we present the RF amplitude spectra measured along the detuning direction, along the yellow arrow indicated in Fig.~\ref{figS4}(b). We use the lever arm $\alpha_\text{L}$ to convert gate voltage into the detuning energy. Here, we use the theoretical model presented in Ref.~\citealp{ranni2023} to reproduce the reflection amplitude spectra as a function of level detuning using the following relation: 

\begin{equation}\label{eqS1}
    |r(\omega)|=|\kappa_c A(\omega)-1|,
\end{equation}
with
\begin{equation}\label{eqS2}
        A(\omega)=\frac{\Tilde{\Gamma}/2-i(\omega-\omega_\text{r})}{[\kappa/2-i(\omega-\omega_\text{r})][\Tilde{\Gamma}/2-i(\omega-E_r/\hbar)]+g^2}
\end{equation}

Figure~\ref{figS4}(d) shows the theoretical results obtained using equation~(\ref{eqS2}) with the cavity DQD coupling constant $g$, tunnel coupling $t_\text{L,R}$, decoherence rate $\Tilde{\Gamma}$ as the fitting parameters. In Figs.~\ref{figS4}(e-g), we compare the measured data and theory results along the A, B, and C line-cuts, marked in Fig.~\ref{figS4}(c). The fitting yields $g$ = $2\pi\:\times\:$$40$~MHz, $t_\text{L,R}$ = $12.2~\mu$eV, and total decoherence rate $\Tilde{\Gamma}$ = $2\pi\:\times\:$$1.1$~GHz. Consequently, the bare cavity-DQD coupling constant $g_0$ is determined to be $2\pi\:\times\:$$45$~MHz. Based on the scatter of the data points and their correspondence to the fits, we estimate an uncertainty of approximately 10\:\% in the estimated parameters. Manually adjusting the parameter values beyond this 10\:\% significantly distorts the fitted results. 

\subsection{The $P(E)$ theory for the single photon process.}

Considering the Gaussian broadening of the DQD levels, the tunnel rate at the single photon absorption limit can be expressed using equation~(2) from the main text:

\begin{equation}\label{pe1}
\Gamma_{\text{L}\rightarrow\text{R}}(E=\hbar\omega_\text{r})=\frac{2\pi}{\hbar}t_\text{L,R}^2\left[J_1^2(2\sqrt{\rho_\text{cav}}\alpha_0)\right]\frac{1}{\sqrt{2\pi\sigma_\text{E}^2}}.
\end{equation}

In the linear response regime, the photon assisted tunnel rate can be described by Taylor expanding equation~(\ref{pe1}) and restricting the series to the first-order term,

\begin{equation}\label{pe2}
\Gamma_{\text{L}\rightarrow\text{R}}(E=\hbar\omega_\text{r}) = \frac{\sqrt{2\pi}}{\hbar\sigma_\text{E}}t_\text{L,R}^2\rho_\text{cav}\alpha_0^2.
\end{equation}

Here, the strength of the zero-point voltage fluctuation $\rho_\text{cav}=\xi^2\cdot\pi Z_0/R_Q$ and $\alpha_0^2=n_\text{c}=2C_\text{r}Q_\text{L}^2Z_0P_\text{in}/Q_\text{ex}\hbar\omega_\text{r}$. Simplifying further:

\begin{equation}\label{pe3}
\Gamma_{\text{L}\rightarrow\text{R}}(E=\hbar\omega_\text{r}) = \frac{\sqrt{2\pi}}{\hbar\sigma_\text{E}}t_\text{L,R}^2 \left[\frac{\xi^2\cdot\pi Z_0}{R_Q}\right] \frac{2C_\text{r}Q_\text{L}^2 Z_0}{Q_\text{ex}}\dot{N}.
\end{equation}

Additional simplification involves the loaded quality factor $Q_\text{L}=\omega_\text{r}/\kappa$, external quality factor $Q_\text{ex}=\omega_\text{r}/\kappa_c$, impedance of the resonator $Z_0=\sqrt{\pi^2L_\text{r}/4C_\text{r}}$, and the resonance frequency $\omega_\text{r}=1/\sqrt{L_\text{r}C_\text{r}}$, where $L_\text{r}$ is the lumped element inductance of the resonator line~\cite{goppl2008}. This yields,

\begin{equation}\label{pe4}
\eta=\frac{\Gamma_{\text{L}\rightarrow\text{R}}(E=\hbar\omega_\text{r})}{\dot{N}} = \frac{\sqrt{2\pi}\cdot\pi}{\hbar\sigma_\text{E}}t_\text{L,R}^2 \left[\frac{\xi^2\cdot\pi Z_0}{R_Q}\right] \frac{\kappa_c}{\kappa^2}.
\end{equation}

Now, we define the square of the bare cavity-DQD coupling constant $g_0^2=(\pi \sqrt{\pi}/32\sqrt{2})\cdot(\omega_\text{r}^2\xi^2\cdot2 Z_0/R_Q)$~\cite{childress2004} and the decoherence rate $\Tilde{\Gamma}=\sigma_\text{E}/h$. With $\xi=0.26$ obtained using the $P(E)$ theory, see the main text, the above relation can be used to calculate $g_0$ = $2\pi\:\times\:$40~MHz. Next, using equation~(\ref{pe4}) we obtain:

\begin{equation}\label{pe5}
\eta=\frac{16}{\Tilde{\Gamma}}\left[\frac{2t_\text{L,R}}{\hbar\omega_\text{r}}g_0\right]^2\cdot\frac{\kappa_c}{\kappa^2}.
\end{equation}

With the cavity-DQD coupling constant $g = g_0\cdot2t_{L,R}/\hbar\omega_\text{r}$ and the cavity photon dissipation rate due to the DQD absorption, $\kappa_\text{DQD}=4g^2/\Tilde{\Gamma}$, equation~(\ref{pe5}) yields:

\begin{equation}\label{pe6}
\eta = \frac{4 \kappa_\text{DQD}\kappa_c}{\kappa^2}.
\end{equation}

Note that, when both $\kappa_\text{in}$ and $\kappa_\text{DQD}$ are small, equation~(\ref{pe6}), derived in the $P(E)$ theory framework, aligns exactly with equation~(4) derived in the Jaynes-Cummings input-output theory limit.

\bibliography{master}